\documentclass[%
 reprint,
 amsmath,amssymb,
 aps,
]{revtex4-1}
\usepackage{epstopdf}
\usepackage{graphics}
\usepackage{graphicx}
\usepackage{dcolumn}
\usepackage{bm}

\newcommand\lsim{\mathrel{\rlap{\lower4pt\hbox{\hskip1pt$\sim$}}
    \raise1pt\hbox{$<$}}}
\newcommand\gsim{\mathrel{\rlap{\lower4pt\hbox{\hskip1pt$\sim$}}
    \raise1pt\hbox{$>$}}} 
    \newcommand{\dm}{\mathrm {dm}}
\newcommand{\br}{\mathrm {b}}  
\newcommand{\vbc}{\mathrm{vbc}}
\newcommand{\bc}{\mathrm{bc}}

\newcommand{\apjs}{{\em ApJS}}
\newcommand{\apjl}{{\em ApJl}}
\newcommand{\physrep}{{\em PhyRev}}
\newcommand{\mnras}{{\em MNRAS}}

\newcommand{\araa}{{\em ARAA}}
\newcommand{\aap}{{\em AAP}}
\newcommand{\ssr}{{\em SSR}}

\usepackage{color}

\begin{document}

\preprint{APS/123-QED}

\title{Generation of Primordial Magnetic Fields on Linear Over-density Scales}

\author{Smadar Naoz}
\homepage{Einstein Fellow} \email{snaoz@cfa.harvard.edu}
\author{Ramesh Narayan}%
\affiliation{%
Institute for Theory and Computation, Harvard-Smithsonian Center for Astrophysics,\\ 60 Garden St.; Cambridge, MA, USA 
}%

%


\date{\today}

\begin{abstract}
Magnetic fields appear to be present in all galaxies and galaxy
clusters. Recent measurements indicate that a weak magnetic
field may be present even in the smooth low density intergalactic medium.
One explanation for these observations is that a seed magnetic field was generated by
some unknown mechanism early in the life of the Universe, and was later amplified
by various dynamos in nonlinear objects like galaxies and clusters.
We show that a primordial magnetic field is expected to be generated in the early
Universe on purely linear scales through vorticity induced by scale-dependent temperature fluctuations or equivalently, a
spatially varying speed of sound of the gas.  Residual free electrons
left over after recombination tap into this vorticity to generate 
magnetic field via the Biermann battery process. Although the battery
operates even in the absence of any relative velocity between dark
matter and gas at the time of recombination, the presence of such a
relative velocity modifies the predicted spatial power spectrum of the
magnetic field.  At redshifts of order a few tens, we estimate a root
mean square field strength of order $10^{-25} -10^{-24}$\,G on
comoving scales $\sim10$\,kpc. This field, which is generated purely
from linear perturbations, is expected to be amplified significantly
after reionization, and to be further boosted by dynamo processes
during nonlinear structure formation.
\end{abstract}

\maketitle

\section{Introduction}
Galaxies in the local Universe have coherent magnetic fields with
strength $\sim10^{-6}$\,G~\cite{Beck+96,Widrow02,vellee04}. Similar
fields strengths are seen in galaxies up to redshift $\sim 2$
~\cite{Widrow02,Bernet+08}. In some cases, the field appears to be even
stronger, e.g., a recent measurement of $30\,\mu$G in star forming
galaxies~\cite{Syan+12}.  One explanation is that the  observed fields 
originated from primordial magnetic fields which were created in the
very early Universe and were later amplified during the formation of
the galaxies. Another possibility is that there were no primordial  fields  and the observed fields were generated spontaneously  during the gravitational collapse of galaxies ~\cite{Kronberg94,DN13}.

There is independent evidence for a pre-galactic seed magnetic field
in the inter galactic medium (IGM). This is based on the lack of
detection of inverse Compton GeV radiation from charged secondaries
associated with extragalactic TeV sources. A magnetic field greater
than $\sim10^{-16}$\,G can deflect  secondaries sufficiently to
explain the observations~\cite{NV10,Ale+10}; the required field
strength has been reduced to $10^{-18}$\,G in a recent study
\cite{Dermer+11}. This evidence for magnetic fields in the IGM
emphasizes the notion that the fields are primordial (see for
  further discussion Ref.~\cite{Ryu+12}), although it is possible that the  fields originated by baryonic outflows from already formed galaxies~\cite{Kronberg94,DN13}.  We note that the absence of
secondary radiation from TeV sources may have nothing to do with a
magnetic field but be the result of beam instabilities which slow down
the particles before they can produce significant inverse Compton
radiation~\cite{broderick12} (but see Ref.~\cite{MB12}). Other recent studies which have
considered the influence of primordial magnetic fields on the cosmic
microwave background (CMB) and Ly$\alpha$ clouds
\cite{Shaw+12,kah+12,PS13} give an upper limit on the present day
large scale magnetic field in the IGM (extended up to $z\sim 5$) of
$\sim 10^{-9}$\,G.
 
In an influential study, Biermann (1950;~Ref.~\cite{Biermann}) showed that currents must
flow whenever a plasma  has a rotational vortex--like
motion. These currents will lead to the generation of magnetic field
starting from zero field. The process has been coined in the
literature as the ``Biermann battery", and several astrophysical
applications have been discussed.
These range from the generation of
magnetic fields in stars~\cite{Biermann,MR62} to seed
magnetic fields on galactic scales~\cite{sub+94,Nick+00,Xu+08,Sub08,Widrow+12}.  The latter
studies typically use nonlinear gas-dynamical processes such as those
that occur in shocks  during structure formation.  

 It has been argued that magnetic fields, at the time of recombination, may be generated on large scales ($>600$~kpc) through second-order couplings between
photons and electrons~\cite{Ichiki+06}. Here we consider smaller scales, and we show that seed magnetic fields can be produced in the early
Universe starting from zero field purely as a consequence of the growth of {\em linear}
over-densities.  We consider the evolution of density and temperature
fluctuations of the baryonic matter after the time of
recombination. We follow the approach described in Ref.~\cite{NB05}, where
the key new effect that permits the generation of magnetic fields is a
spatially varying speed of sound (see below). We also consider the
effect of the relative velocities between the dark matter and baryons
at the time of recombination~\cite{Tes+10a}. The latter effect has
been shown to have a considerable effect on the evolution of over
densities at high redshifts
\cite{Tes+10a,Tes+10b,Stacy+10,Greif10,Visbal+12,Naoz+11a,OLMc12,Naoz+12}. Here we show
that it has a noticeable effect also on the growth of the magnetic
field. 

Throughout this paper, we adopt the following cosmological parameters:
($\Omega_\Lambda$, $\Omega_{\rm M}$, $\Omega_b$, n, $\sigma_8$,
$H_0$)= (0.72, 0.28, 0.046, 1, 0.82, 70 km s$^{-1}$ Mpc$^{-1}$)
\cite{wmap5}.

\section{Linear evolution of  over densities in the early Universe}\label{sec:Lin}

After cosmic recombination, the baryonic gas in the Universe decouples
mechanically from the photons, but remains thermally coupled down to
$z \sim 150$. This coupling is a result of CMB photons scattering off
the residual free electrons, which constitute a fraction $\sim10^{-4}$
of the bound electrons.  Even at $z<150$ the baryons still retain some
memory of this heating, which induces scale-dependent temperature fluctuations or equivalently, a spatially varying speed of
sound in the gas. Naoz \& Barkana (2005, Ref.~\cite{NB05}) took this effect into account and
computed the linear growth of baryonic density and temperature
fluctuations separately~\footnote{ The effect on the dark
  matter evolution is much more modest since the baryons contribute
  only a small fraction of the total gravitational acceleration felt
  by the dark matter.}.  At large wavenumbers ($k > 100$ Mpc$^{-1}$)
the growth of baryon density fluctuations is changed significantly by
the effect of the inhomogeneous sound speed, by up to $30\%$ at
$z=100$ and $10\%$ at $z=20$. This has an important impact on high-$z$
gas rich halos~\cite{NB07}.

It was shown recently that not only  the amplitudes of
the dark matter and baryonic density fluctuations are different at early
times, so too are their velocities~\cite{Tes+10a}. After recombination, the sound
speed of the baryons drops dramatically, while the dark matter
velocity remains high. As a result, the relative velocity of baryons
with respect to the dark matter becomes supersonic.
This relative velocity, which is generally referred to as
the ``stream velocity'' in the literature, remains coherent on scales
of a few Mpc and is of the order of $\sim 30~{\rm km\,s^{-1}}$ at the
time of recombination~~\cite{Tes+10a}.


%
\begin{figure}[!t]
\includegraphics[width=\linewidth]{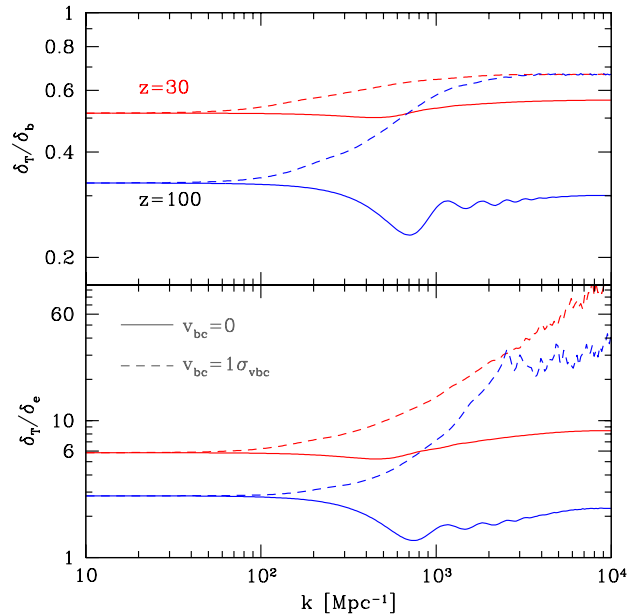}
\caption{ 
Perturbation ratios $\delta_T /\delta_b$ (top panel) and
  $\delta_T / \delta_e$ (bottom panel) as a function of wavenumber
  $k$. We consider two cases: no stream velocity, $v_{\rm bc}=0$
  (solid lines), and a typical stream velocity, $v_\bc=1\sigma_\vbc$
  (dashed lines). Results are shown for two redshifts, $z=100$ (blue
  lines) and z=30 (red lines).}
   \label{fig:dTde} 
\end{figure}

For completeness we write here the coupled second order differential
equations that govern the evolution of the dimensionless density
fluctuations of the dark matter $\delta_{\dm}$ and of the baryons
$\delta_{\br}$:
\begin{eqnarray}\label{g_T}
\ddot{\delta}_{\dm} + 2H \dot {\delta}_{\dm}- f_{\dm}\frac{2 i}{a}  {\bf v}_{\bc} \cdot {\bf k} \dot\delta_{\dm}  & = &
 \\
\frac{3}{2}H_0^2\frac{\Omega_{m}}{a^3}
\left(f_{\br} \delta_{\br} + f_{\dm} \delta_{\dm}\right)& +& \left( \frac{  {\bf v}_{\bc} \cdot {\bf k}} {a} \right)^2 \delta_{\dm}   \nonumber  \\
\ddot{\delta}_{\br}+ 2H \dot {\delta}_{\br}  \ \ \ \ \ \ \ \ \ & = &
  \\
\frac{3}{2}H_0^2\frac{\Omega_{m}}{a^3} \left(f_{\br}
\delta_{\br} + f_{\dm}
\delta_{\dm}\right) &-& \frac{k^2}{a^2}\frac{k_B\bar{T}}{\mu}
\left(\delta_{\br}+\delta_{T}\right) \ , \nonumber \label{eq:db}
 \end{eqnarray}
where $\Omega_m$ is the present day matter density as a fraction of
the critical density, $k$ is the comoving wavenumber of the
perturbation, ${\bf v}_{\bc}$ is the relative velocity between the
baryons and dark matter in a local patch of the Universe, $a$ is the
scale factor of the Universe, $H_0$ is the present day value of the Hubble parameter, $\mu$ is the mean molecular weight of
the gas, 
$\bar{T}$ is the mean temperature of the baryons, $f_{\br}$ ($f_{\dm}$) is the cosmic baryonic (dark matter) fraction and $\delta_T$ is
the dimensionless fluctuation in the baryon temperature. Derivatives are with respect to the clock time. These equations are
a compact form of equations (5) in Ref.~\cite{Tes+10a}, where we have used
the fact that $v_{\bc} \propto 1/a$, and have included the pressure term
appropriate to the equation of state of an ideal gas
e.g.,~\cite{NB05}. 

The linear evolution of the baryon temperature fluctuations may be
written down similarly~\cite{BL05,NB05}. Including an additional term
due to fluctuations of the electron over density $\delta_e$:
\begin{eqnarray}
\label{gamma} \dot{\delta}_T &=& \frac{2}{3} \dot{\delta}_\br + \frac{x_e(t)} {t_{\gamma}}a^{-4} \bigg \{
\frac{\bar{T}_{\gamma}} {\bar{T}} \left(\delta_{T_{\gamma}} -\delta_T
\right)\nonumber \\ 
&+&
\left(\delta_{\gamma}+\delta_e\right)\left( \frac{\bar{T}_{\gamma}}{\bar{T}} -1\right)
 \bigg\}\ ,
\end{eqnarray}
where $\delta_{\gamma}$ is the photon density fluctuation,
$t_{\gamma}^{-1}=8.55 \times 10^{-13}\,{\mathrm{yr}}^{-1}$, and $\bar{T}_{\gamma}$ and
$\delta_{T_{\gamma}}$ are the mean photon temperature and its
fluctuation, respectively.  

The evolution of the mean free electron fraction $x_e$ as a function
of time is 
\begin{equation}
\dot{x}_e= - \alpha_B(T) x_e^2 n_H (1+y) \ ,
\end{equation}
where $ \alpha_B(T)$ is the case B recombination coefficient as a
function of the gas temperature, $n_H$ is the {\it total} hydrogen
number density, and $y=n_{He}/n_H$ where $n_{He}$ is the helium number
density. Fluctuations in the electron density, $\delta_e=\Delta n_e/n_e=\Delta x_e/x_e$, evolve
according as
\begin{equation}
\dot{\delta}_e= - \alpha_B(T) (1+y) x_e n_H(\delta_b + \delta_e)  \ .
\end{equation}
We show below  that the magnetic field grows
because of the presence of the residual free electrons. It is highly
sensitive to the evolution of their fractional fluctuations
$\delta_e$, but not to the  actual electron number density.

Equation~(\ref{gamma}) describes the
evolution of the gas temperature in the post-recombination era but
before the formation of the first galaxies, when the only external
heating arises from Compton scattering of the remaining free electrons
on the CMB photons~~\cite{NB05}. The first term in Equation~(\ref{gamma}) describes
the adiabatic cooling of the gas, while the second term is the result
of Compton interactions. An important effect of this equation is that
it introduces a scale dependent behavior in the fluctuations of the
temperature, free electron density and baryon density. In this full
thermal evolution calculation, the sound speed ($c_s^2=dp/d\rho$,
where $p$ is the pressure of the gas), varies spatially, simply
because $\delta_{\br}$ and $\delta_T$ have the following relation\begin{equation}
1+\frac{\delta_T}{\delta_{\br}}=\frac{c_s^2}{k_B \bar{T}/\mu}=\gamma_{\rm eff} \ ,
\end{equation}
where $\gamma_{\rm eff}$ is a scale dependent, effective adiabatic index.

In Figure \ref{fig:dTde} we show the ratios $\delta_T /\delta_b$ (top
panel) and $\delta_T / \delta_e$ (bottom panel) as a function of $k$.
At the largest scales (smallest $k$), the baryons follow the dark
matter density, and $\delta_T /\delta_b$ evolves from $1/3$ (at high
redshift where the baryons are tightly coupled to the relativistic CMB)
to $\sim 2/3$ (lower redshift where the baryons expand adiabatically as
an independent nonrelativistic fluid). Considering first the  zero stream velocity case,  small scales (large $k$) at high redshift show  Jeans scale oscillations which are suppressed at lower redshift (there is only  a slight minimum for $z=30$). For $v_{\bc}=1\sigma_{\vbc}$, the small scale baryon fluctuations drop, and are less important compared to the Compton heating [see Equation (\ref{gamma})] which results in a slight increase of the temperature fluctuations (compared to the zero stream velocity). These two  effects result in an increase of the ratio $\delta_T /\delta_b$  as a function of scale.   The free electron fluctuations are further suppressed in the case of $v_\bc=1\sigma_{\vbc}$ compared to the  case of zero stream velocity which results in a larger increase  in the ratio  $\delta_T / \delta_e$.

\section{Biermann Battery in an Expanding Universe}\label{sec:Bier}

The evolution of the magnetic field via the Biermann battery process
is described by a simple combination of the Maxwell--Faraday equation
and the generalized Ohm's law e.g., Ref.~\cite{Kulsrud+05}. Since we
are interested in magnetic field evolution over cosmic times, we need
to work with the Biermann battery equation in a flat expanding
Universe.  In this case, we find that the differential equation
for the clock time evolution of the magnetic field ${\bf B}$ is given
by:
    \begin{equation} \label{eq:Bier}
   \frac{\partial}{\partial t}\left( a^2 {\bf B}\right)=a \nabla \times \left( {\bf u} \times {\bf B}\right) -  c \frac{\nabla n_e \times \nabla P_e}{e n_e^2} \ ,
      \end{equation}
where $n_e$ and $P_e$ are the electron number density and pressure
respectively, $e$ is the electron charge, ${\bf u}$ is the peculiar
velocity of the gas, and the spatial derivatives are with respect to
co-moving coordinates. To relate to the literature \cite{Dettmann+93,Subramanian10} this equation can be reduced to the familiar form of the Biermann battery by rescaling $\tilde{\bf B}=a^2 {\bf B}$, $\tilde{n}_e=n_e/a^3$, and $\tilde{P}_e=P_e/a^4$, for conformal time ($\eta$) where $\partial / \partial \eta = a \partial / \partial t$. The resulting equations are those used for example in describing a recent laboratory experiment of the Biermann battery \cite{Gregori+12}. Below we do not  rescale the equations since the temperature and density fluctuations of the gas have a more complicated dependence on the scale factor \cite{NB05}.   The term $ \nabla \times \left( {\bf u} \times
{\bf B}\right)$, in equation (\ref{eq:Bier}), describes flux freezing, i.e., the magnetic flux
through any closed contour embedded in the plasma is conserved under
plasma motions.  The last term is the Biermann battery term.  This
term is proportional to the derivative with respect to time of the
vorticity of the electrons; a vortex--like motion of the electrons
produces an rotational electric field, and through this a magnetic field~\footnote{Note that here we considered vorticity in the gas, for possible  contribution of vorticity in the dark matter fluid near the time of equality to structure formation was considered in \cite{MT94}.}.

\begin{figure}[htb]
\includegraphics[clip,scale=0.45]{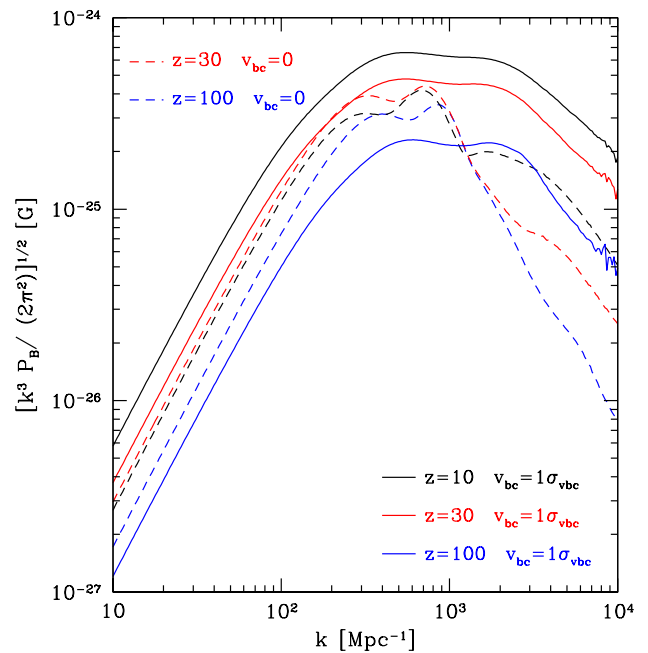}
\caption{Root mean square magnetic field generated by the Biermann
  battery as a function of wavenumber. Two cases are shown: no stream
  velocity, $v_{\bc}=0$ (solid lines), and a typical stream velocity,
  $v_{\bc}=1\sigma_{\vbc} $ (dashed lines).  Three redshifts are
  considered: $z=100$ (blue lines), $z=30$ (red lines), $z=10$ (black  lines). }
\label{fig:Bofk}
\end{figure}

Consider now the Biermann term $ c \nabla n_e \times \nabla P_e / e
n_e^2$.  The electron pressure is given by $P_e=n_ek_B T$, where,
following~\cite{NB05}, we have set $T_e=T$. Expanding the relevant
quantities to linear order, i.e., $n_e=\bar{n}_e(1+\delta_e)$ and
$T=\bar{T}(1+\delta_T)$, and neglecting the flux-freezing
term~\footnote{ When approximating the peculiar velocity of the electrons  to be $c_s^2\delta_e k/(H a)$, we find that the for $10^{-24}$~G 
the flux freezing term is still much
smaller then the Biermann term (between  $4-6$ orders of magnitude), and thus we neglect it for the future
calculation~\cite[e.g.,][]{Kulsrud+05}.}, equation (\ref{eq:Bier}) can be written as:
 \begin{equation}\label{eq:Bier2}
\frac{\partial}{\partial t}\left( a^2 {\bf B}\right)= -  \frac{ c k_B\bar{T}}{e}\nabla \delta_e \times \nabla \delta_T \ .
\end{equation}
Note that the number density of free electrons cancels out.  Thus the
Biermann effect depends only on the fluctuations in the fractional
electron density $\delta_e$ and not on the actual density $n_e$
itself. Therefore, the fact that the ionization fraction of the gas is
very low ($\sim10^{-4}$) is not important.

The right hand side of equation (\ref{eq:Bier2}) may be written in
Fourier space as
 \begin{eqnarray} 
  -\nabla \delta_e \times \nabla \delta_T &=&\frac{1}{2}\int{\frac{d^3k_1}{(2\pi)^3}}\frac{d^3k_2}{(2\pi)^3} \left({\bf k}_1\times {\bf k}_2\right) e^{i{\bf r}({\bf k}_1+{\bf k}_2)} \nonumber \\
 &\times& \bigg [ \delta_e({\bf k}_1)\delta_T({\bf k}_2) -  \delta_e({\bf k}_2)\delta_T({\bf k}_1) \bigg ] \ . \nonumber \\ 
\end{eqnarray} 
Fourier transforming both sides of equation (\ref{eq:Bier2}), we then
find
  \begin{eqnarray}\label{eq:temp1} 
  \frac{\partial}{\partial t}\left( a^2 {\bf B_k}\right)&=&\frac{1}{2} \frac{ c k_B\bar{T}}{e} \int{\frac{d^3k_1}{(2\pi)^3}} \left( {\bf k}_1\times [{\bf k}-{\bf k}_1]\right) \\
 &\times& \bigg [ \delta_e({\bf k}_1)\delta_T({\bf k}-{\bf k}_1) -  \delta_e({\bf k}-{\bf k}_1)\delta_T({\bf k}_1) \bigg ] \ , \nonumber 
 \end{eqnarray}
where ${\bf B_k}$ has units of G\,Mpc$^3$.  Note that the over
densities that appear in the above equation are complex, i.e.,
$\delta(k)=|\delta(k)| e^{i\phi_k}$, where each $\phi_k$ represents a
random phase which is uniformly distributed over the interval 0 to
$2\pi$. The phases disappear below when we finally compute the power
spectrum of the magnetic field.

Before proceeding, we note that the Biermann battery produces a
magnetic field only if the gradients $\nabla \delta_e$ and $\nabla
\delta_T$ in equation (\ref{eq:Bier2}) are not parallel to each
other. The equivalent condition in Fourier space is that the quantity
in square brackets in equation (\ref{eq:temp1}) should be
non-vanishing.  The latter condition requires  the ratio
$\delta_T({\bf k})/\delta_e({\bf k})$  to vary with scale. This is precisely where the correct
treatment of the gas thermodynamics, as described in Ref.~\cite{NB05}, is
critical. As Figure \ref{fig:dTde} shows, the ratio of temperature to
density fluctuations does vary with $k$, and therefore we expect the
cosmological Biermann battery to operate even within linear
perturbation theory.

Let us define
$\Delta_{e,T}({\bf k,k}_1)=\delta_e({\bf k}_1)\delta_T(|{\bf k}-{\bf k}_1|) -
\delta_e(|{\bf k}-{\bf k}_1|)\delta_T({\bf k}_1) $.
Equation (\ref{eq:temp1}) then becomes
\begin{equation}\label{eq:temp2}
a H\frac{\partial  \left(a^2 {\bf B}_{\bf k}\right)}{\partial a} =\frac{ c k_B}{e}\int{\frac{d^3k_1}{(2\pi)^3}} \bar{T}(t) \left( {\bf k}_1\times {\bf k}\right) \Delta_{e,T}({k,k}_1) \ ,
 \end{equation}
where $\partial / \partial a \equiv a H \partial / \partial t$.  In this equation, only $\Delta_{e,T}$ and $\bar{T}$ depend on
the time $t$ (or equivalently the scale factor $a$).  Thus we can
write equation (\ref{eq:temp2}) as
\begin{equation}
 {\bf B}_{\bf k}(a) =\int{\frac{2\pi dk_1 \sin\theta d\theta}{(2\pi)^3}} \beta(a,k,k_1,\theta) \left( {\bf k}_1\times {\bf k}\right)  \ ,
 \end{equation}
where the quantity $\beta=\beta(a,k,\sqrt{k^2+k_1^2-2kk_1\cos \theta})$ satisfies
 \begin{equation}\label{eq:beta}
a H\frac{ \partial (a^2 \beta(a,{k,k}_1))}{\partial a}= \frac{ c k_B}{e}\bar{T}(a)\Delta_{e,T}({k,k}_1) \ .
\end{equation}
By numerically integrating equation (\ref{eq:beta}), we can calculate
the two dimensional array of values $\beta({k,k}_1)$ as a function of
scale $a$ or redshift $z$. These $\beta$ values still include the
random phases $\phi_k$. However, the phases are eliminated when we
compute the power spectrum of the magnetic field $P_{B}$.
The result is 
 \begin{eqnarray}
P_B \equiv \langle{\bf B_k B^\star_k}\rangle &= & \\
 &&\hspace{-0.3in} \frac{1}{V} \int \frac{2\pi dk_1 \sin\theta d\theta}{(2\pi)^3} | \beta(a,k,k_1,\theta)|^2 (k_1 k \sin\theta)^2  \ ,\nonumber 
\end{eqnarray}
where $V$ is the volume.

In Figure \ref{fig:Bofk} we show the power spectrum of the magnetic
field as a function of wavenumber $k$ for different redshifts. The
quantity $\sqrt{k^3 P_B}$ has units of gauss. Note that the magnetic
field grows most strongly on the Jeans mass scale of the baryons. This
is apparent in the case of zero stream velocity, where the first peak
is around $k^{-1} \sim16$~kpc [comoving] at $z=100$, corresponding to
a mass scale $\sim 7\times 10^4$~M$_\odot$. This mass scale is
  slightly above the minimum mass for which baryonic gravitational
  instabilities can still grow~\cite{NB07,NBM,Naoz+10}. The second
peak, where the power is maximum, is associated with smaller scales
$\sim 7$~kpc [comoving], which correspond to where the most dramatic
variation of the ratio $\delta_T/\delta_e$ occurs (see Figure
\ref{fig:dTde}).  For the case of $v_{\bc}=1\sigma_{\vbc}$, we see the
inverse behavior. Here the first peak (larger scales) has more
power than the second peak (smaller scales).  Note that our use of linear theory is justified, since the density perturbations are still linear for scales smaller than $\sim 1000\,{\rm Mpc^{-1}}$ [comoving]
and become nonlinear only at $z<10$
(see Fig.~6 in Ref.~\cite{BL01}, see also \footnote{Note that annihilation of magnetic fields due to resistivity is important only on much smaller scales ($\sim 4$~AU at $z=100$ and $\sim 400$~AU at $a=10$).}).


\section{Discussion}\label{disc}

We have shown that seed magnetic fields can be produced from zero initial
magnetic field on cosmological linear over density scales through the
Biermann process. The typical field strength is $\sim
10^{-25}-10^{-24}$\,G. These seeds fields may later be amplified via
nonlinear dynamo processes~\cite{sub+04,sur+10} and are
perhaps responsible for the present day magnetic fields in galaxies. Note that baryonic outflows can still  contribute to the IGM magnetic field~\cite{Kronberg94}.
The Biermann battery mechanism requires a vortex like motion in the
plasma. We have demonstrated that the spatially varying speed of sound
of gas in the early Universe produces this vorticity in the residual
free electrons.
The process does not depend on the
fraction of free electrons in the Universe (since the electron number
density cancels in the Biermann term), but only on fluctuations in
this quantity. 

During reionization, the temperature of the baryons as well as 
temperature fluctuations will increase. This will lead to even larger
magnetic fields since equation (\ref{eq:Bier2}) shows that the
magnetic field growth depends linearly on $\bar{T}$, and the
temperature after reionization increases to $\bar{T}\to 10^4$~K. The
temperature and electron fraction fluctuations are also expected to
increase substantially~\cite{PF07}. Thus, the magnetic field
could potentially increase post-reionization by $4-6$ orders of
magnitude, bringing it close to the $10^{-18}$\,G estimated from
observations~\cite{NV10,Ale+10,Dermer+11}. This value is about $6$ orders of magnitude smaller compared to other mechanisms in the literature that operate  on the relevant scales (see Ref.~\cite{DN13} for review), but comparable to Ref.~\cite{MB11}. However, the evolution of
$\delta_e$ and $\delta_T$ during and after reionization is model
dependent. In contrast, we have shown in this paper that, even before
reionization, magnetic field can be generated as part of the linear
growth of perturbations in the Universe, and that the field strength due to this process 
can be estimated robustly with few uncertainties.


The effect described here  (following Ref.~\cite{NB05}) produces a vorticity in the baryonic gas  on the order of  $\sim 10^{-20}$~s$^{-1}$ at $z\sim 10$ on scales $\sim 6$~kpc. During  reionization, as in the case of the magnetic field, the vorticity in the gas may again increase by $4-6$ orders of
magnitude, bringing it close to  $10^{-15}$~s$^{-1}$, which is the  vorticity of the Milky Way Galaxy in the solar neighborhood. 
 

Future measurements of the magnetic field in the IGM and in filaments
(for example via Faraday rotation in the CMB,
  Ref.~\cite{KL96,Xu+06}) would be helpful to further clarify the role of
seed magnetic fields.  Already, lower bounds on the magnetic field in
large scale structures~\cite{NV10,Ale+10,Dermer+11,Ryu+98,Ryu+08,Xu+06} suggest
that there must be a primordial seed field in the Universe. The
Biermann Battery process described here, which operates through a
spatially varying speed of sound, can naturally explain these seeds.
Our calculation suggests that different coherent patches in the
Universe with different stream velocities may have up to an order of
magnitude variation in their magnetic fields.  Thus, seed magnetic
fields could conceivably be used in the future to study the stream
velocity distribution in the Universe.

\begin{acknowledgments}
We thank Antoine Bret, Avi Loeb and Lorenzo Sironi for useful discussions. We also thank Francesco Miniati and the anonymous referees for useful comments on the manuscript.
SN was  supported by NASA
through an Einstein Postdoctoral Fellowship awarded by the Chandra
X-ray Center, which is operated by the Smithsonian Astrophysical
Observatory for NASA under contract PF2-130096.   RN acknowledges partial support from  NASA grant
  NNX11AE16G.
\end{acknowledgments}


\bibliography{cosmo}

\end{document}